\documentclass{ws-procs9x6}            
\begin{document}
\title{Holographic Wigner distributions for the pion}

\author{Mohammad Ahmady$^{a,*}$, Chandan Mondal$^{b,\dagger}$, Ruben Sandapen$^{c,\dagger}$,\\ James P. Vary$^{d,\S}$ and Xingbo Zhao$^{a,e,\parallel}$}

\address{$^a$Mount Allison University, Sackville, New Brunswick, Canada, E4L 1E6\\
$^b$Institute of Modern Physics, Chinese Academy of Sciences, Lanzhou-730000, China\\
$^c$Acadia University, Wolfville, Nova-Scotia, Canada, B4P 2R6\\
$^d$Iowa State University, Ames, IA 50011, USA\\
$^e$School of Nuclear Science and Technology, University of Chinese Academy of
Sciences, Beijing 100049, China\\
$^*$mahmady@mta.ca\\
$^\dagger$mondal@impcas.ac.cn\\
$^\dagger$ruben.sandapen@acadiau.ca\\
$^\S$jvary@iastate.edu\\
$^\parallel$xbzhao@impcas.ac.cn
}









\begin{abstract}
We study the Wigner distributions of the pion using a holographic light-front pion wavefunction with dynamical spin effects to reveal its multidimensional structure.
\end{abstract}


\bodymatter

\section{Introduction}\label{sec1}
Wigner distributions in QCD, commonly known as phase-space distributions, were first introduced by Ji \cite{Ji:2003ak}. 
After appropriate phase-space reductions, these distributions reduce to generalized parton distributions (GPDs)  and transverse momentum dependent parton distributions (TMDs) which are measurable in high energy experiments  (For a review on these distributions and the experiments to measure them, see \cite{Ji:1998pc,Goeke:2001tz,Mulders:1995dh,Boer:1997nt}). GPDs allow us to have a three dimensional picture of the hadron in position space \cite{Burkardt:2002hr}. On the other side, TMDs contains three dimensional information regarding the spin-spin and spin-orbit correlations in momentum space \cite{Lorce:2011zta}. Wigner distributions for spin-$\frac{1}{2}$ systems have been investigated in different models e.g.,  light-cone chiral quark soliton model \cite{Lorce:2011kd}, light-front dressed quark model \cite{Mukherjee:2014nya}, light-cone spectator model\cite{Liu:2015eqa}, AdS/QCD inspired quark-diquark model \cite{Chakrabarti:2016yuw,Chakrabarti:2017teq} as well as light front QED model \cite{Kumar:2017xcm}.

Here, we study  the Wigner distributions  for different quark polarizations in the pion using a holographic light-front pion wavefunction which includes dynamical spin effects.  It has been observed that such effects allow for an excellent simultaneous description of a wide range of data: the decay constant, charge radius, spacelike EM and transition form factors, as well as, after QCD evolution, both the parton distribution function and the parton distribution amplitude data with a  single universal AdS/QCD scale \cite{Ahmady:2018muv}. Recently, this spin-improved holographic wavefunction has been used to predict the leading twist TMDs of the pion \cite{Ahmady:2019yvo}.
\begin{figure}[htp]
\begin{center}
\includegraphics[width=0.44\textwidth]{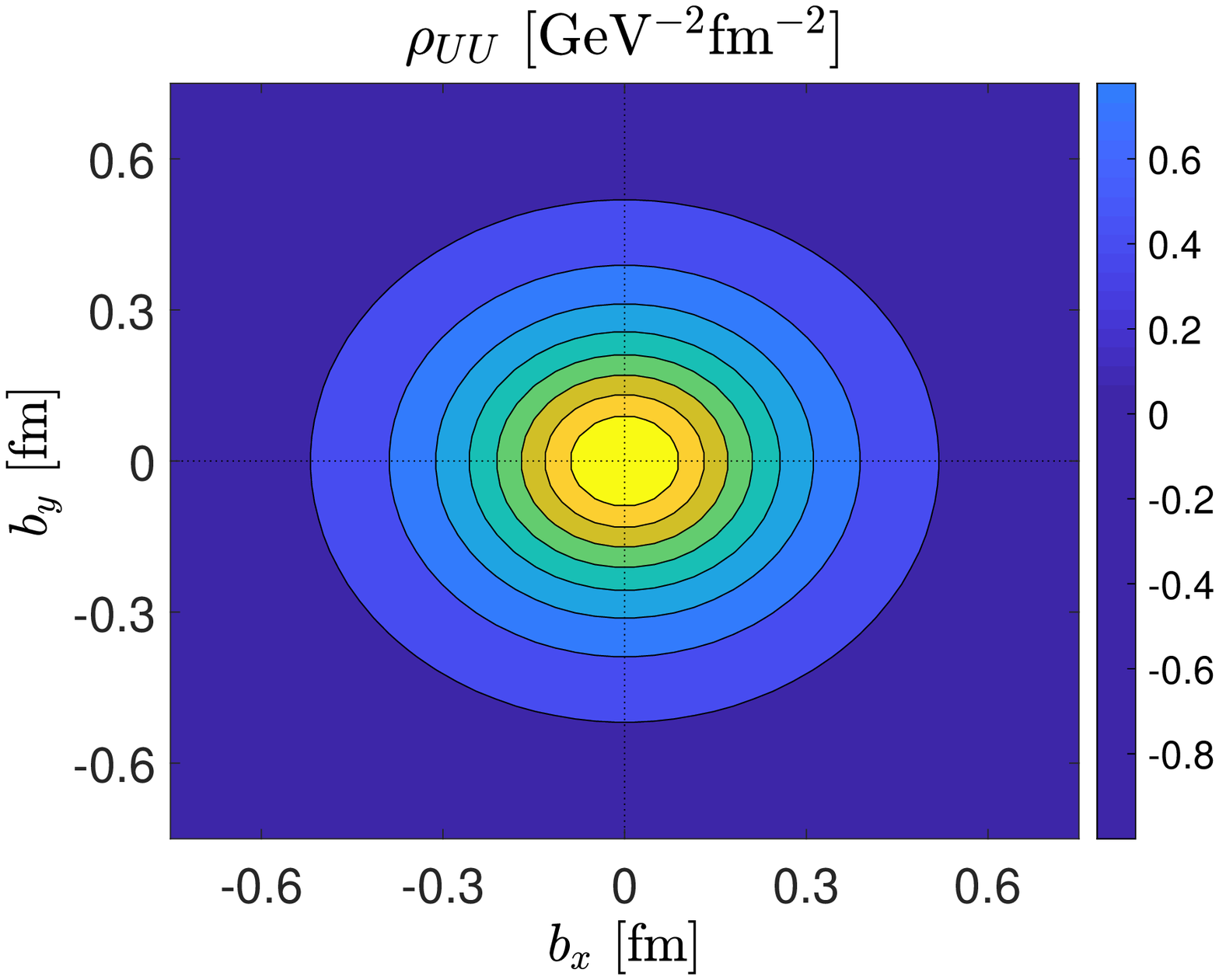}\includegraphics[width=0.44\textwidth]{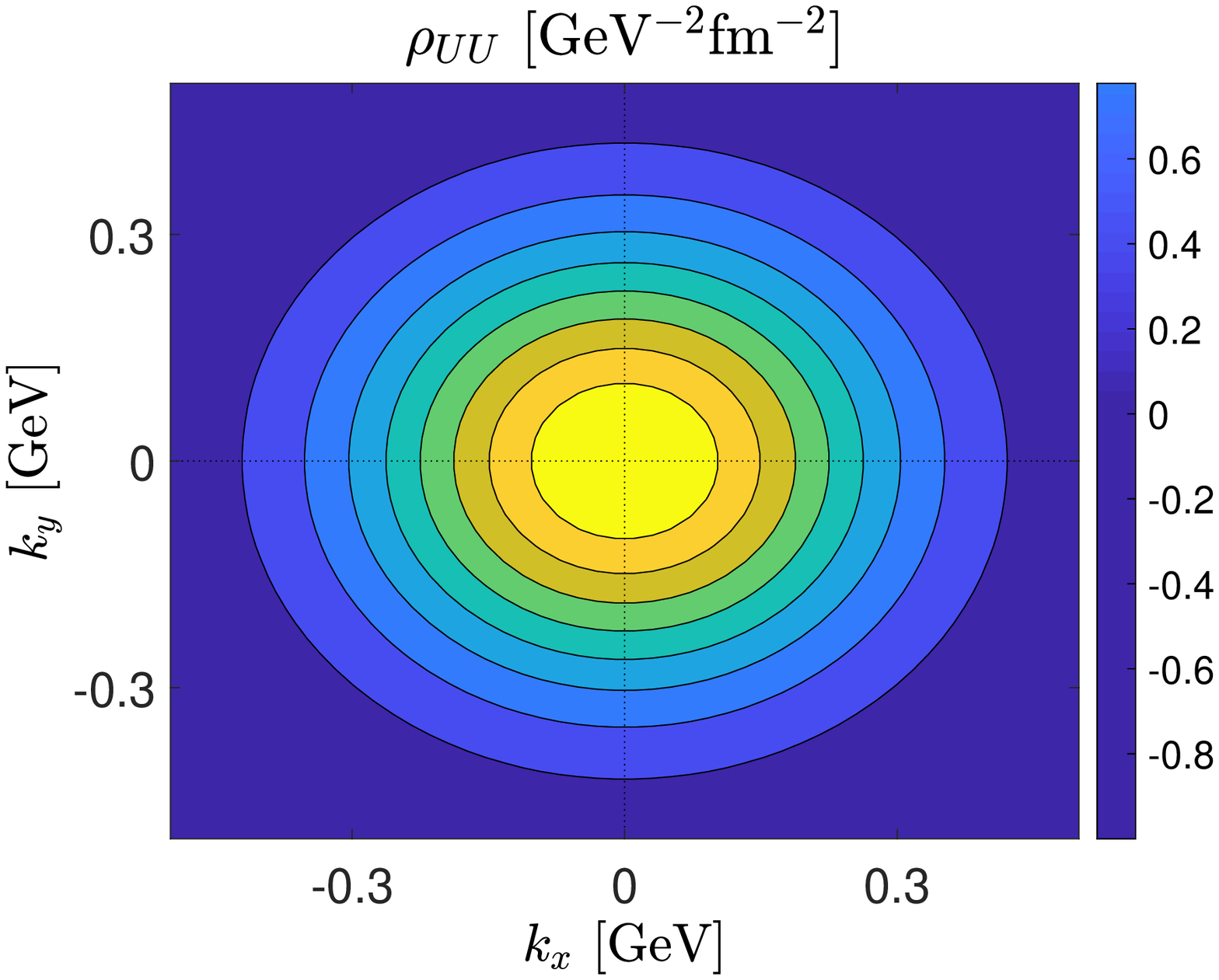}
\includegraphics[width=0.44\textwidth]{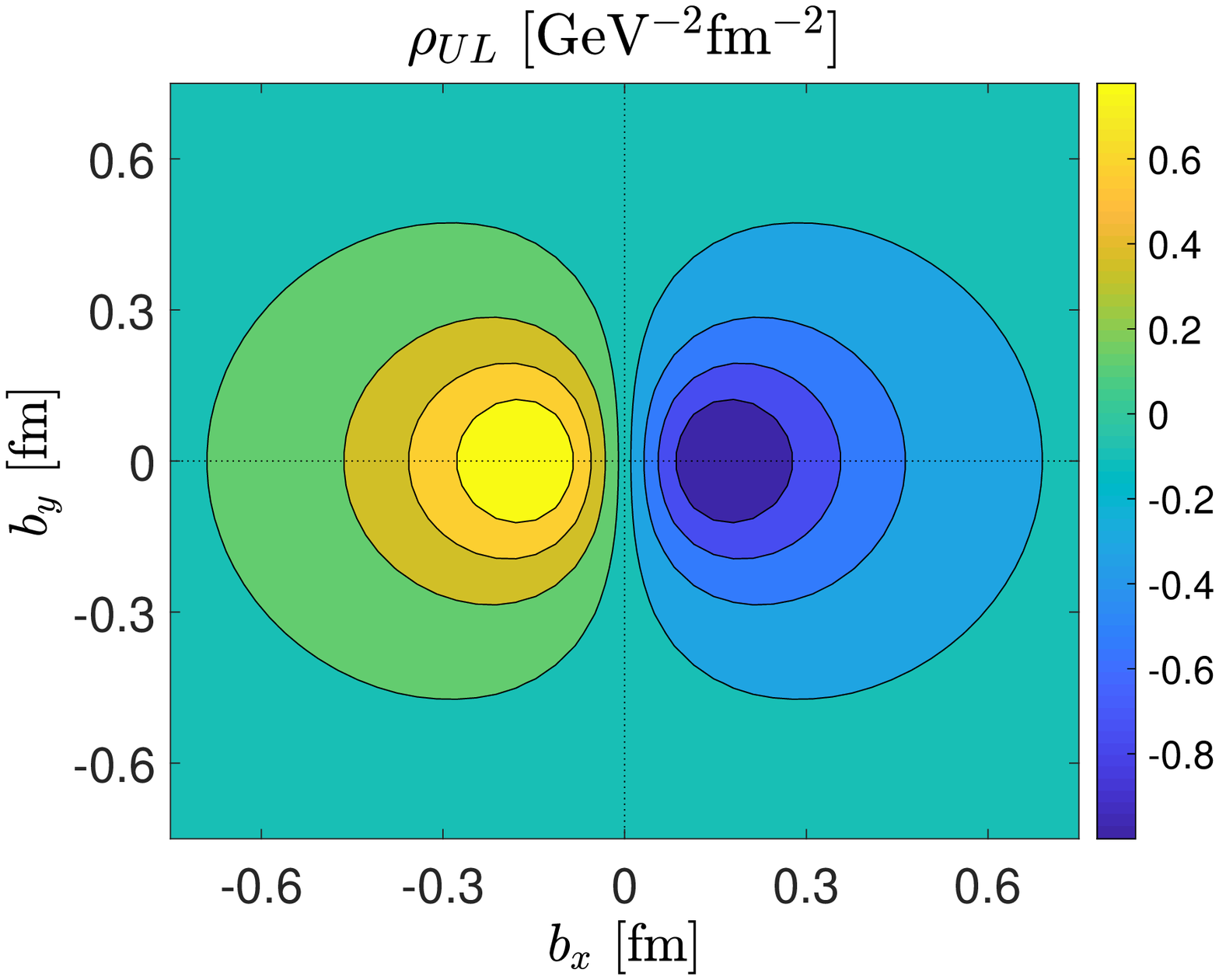}
\includegraphics[width=0.44\textwidth]{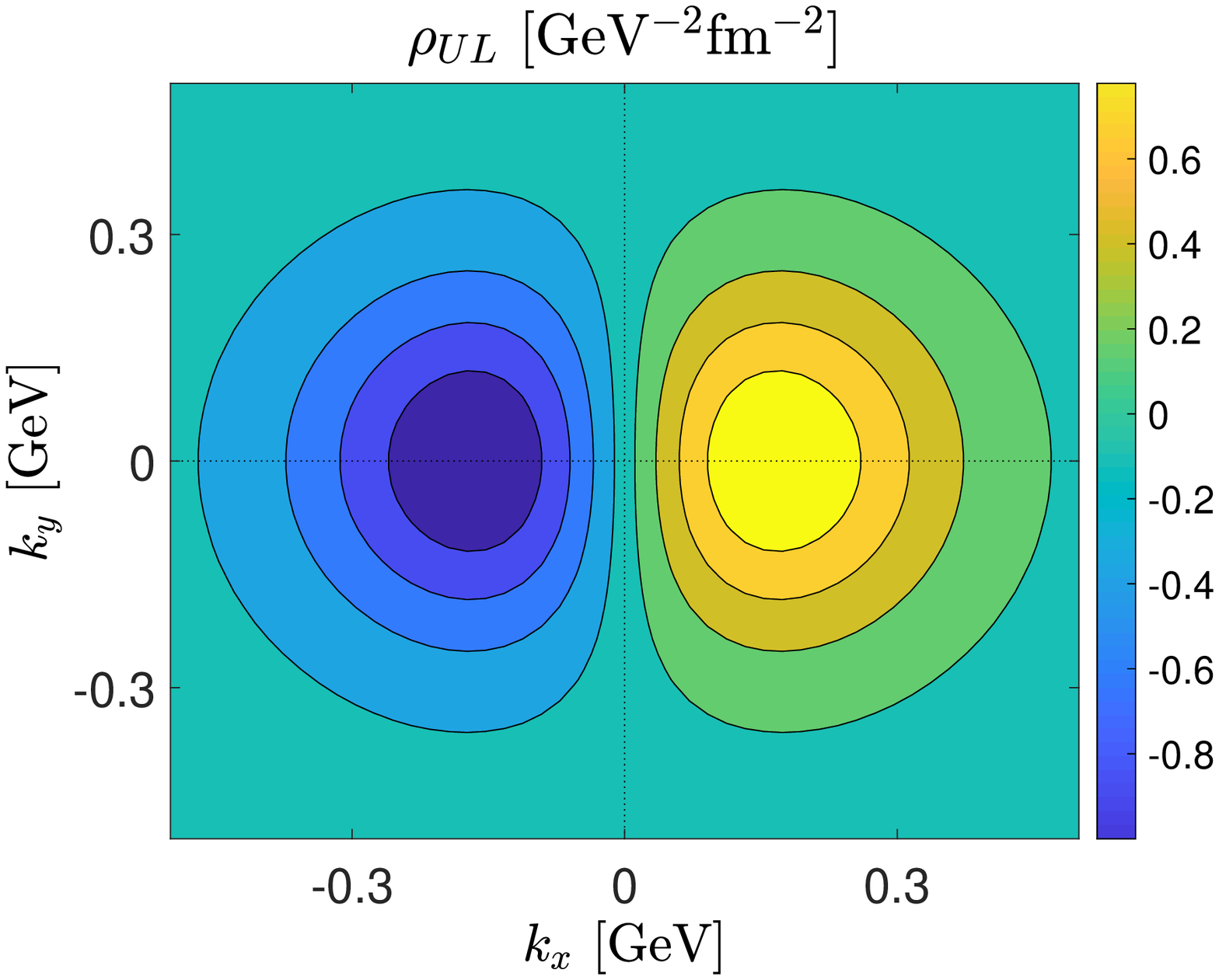}
\includegraphics[width=0.44\textwidth]{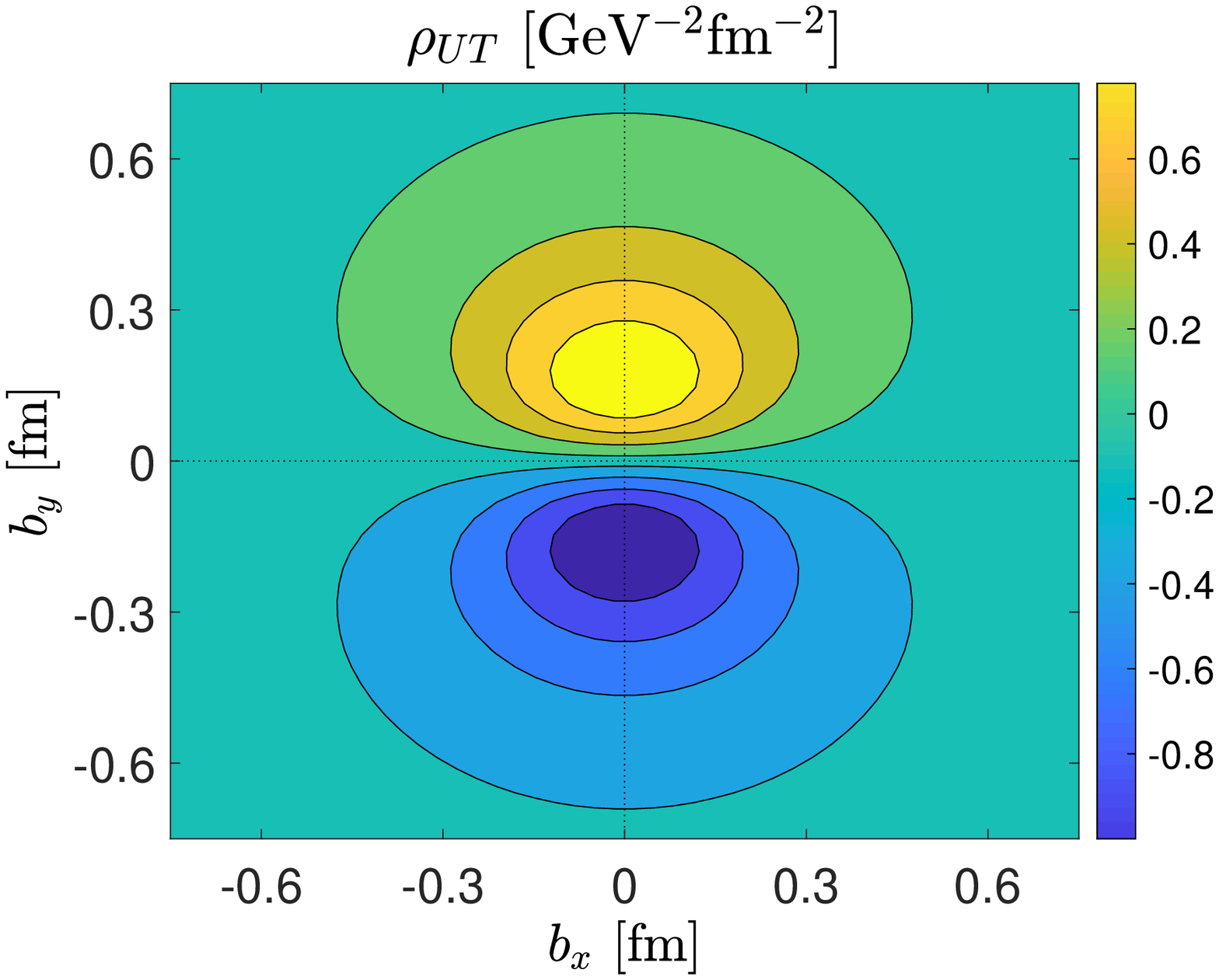}
\includegraphics[width=0.44\textwidth]{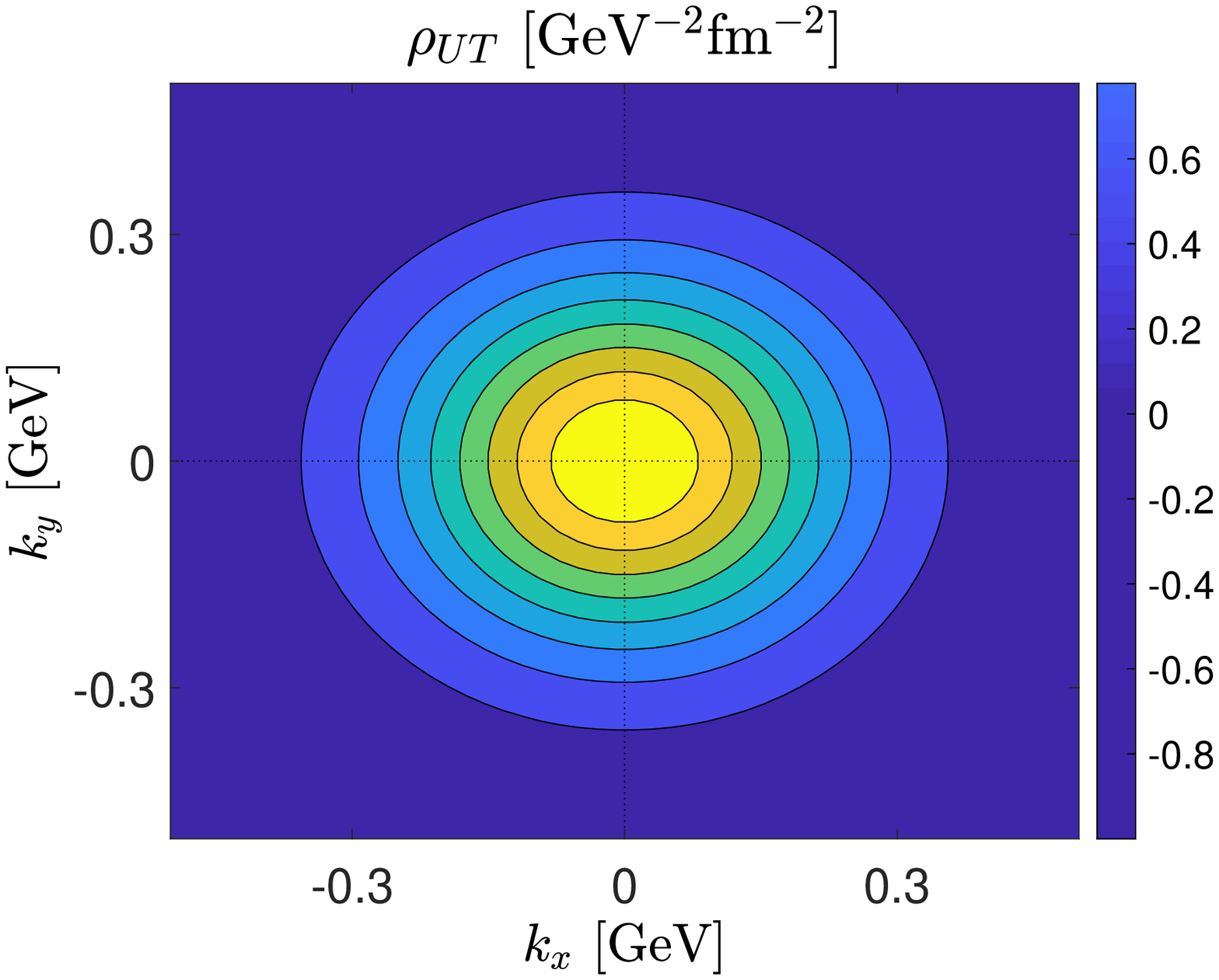}
%
\caption{Wigner distribution $\rho(\bm b_\perp,\bm k_\perp)$ of the unpolarized quark ({\it upper panels}), the longitudinal polarized quark ({\it middle panels}), and  the transversely polarized quark ({\it lower panels}) inside the pion. ({\it Left panels}) the distributions are in the impact-parameter space with fixed transverse momentum $\bm k_\perp = k_\perp \hat{\bm e}_y$ and $k_\perp = 0.3 $ GeV.  ({\it Right panels}) the distributions are in the transverse-momentum space with fixed impact parameter $\bm b_\perp = b_\perp \hat{\bm e}_y$ and $b_\perp = 0.3$ fm. 
}
\label{fig_1}
\end{center}
\end{figure}
\begin{figure}[htp]
\begin{center}
\includegraphics[width=0.44\textwidth]{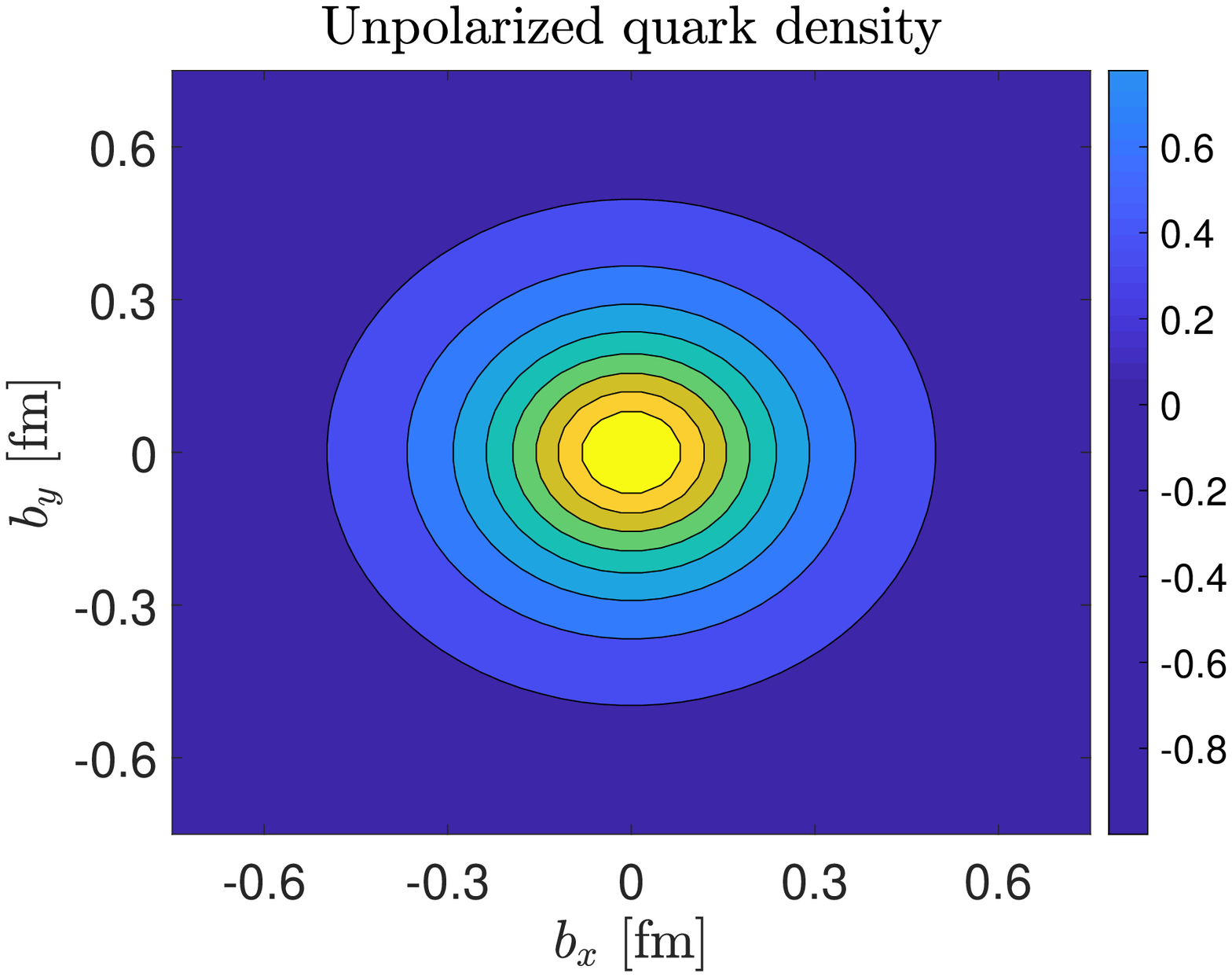}\includegraphics[width=0.44\textwidth]{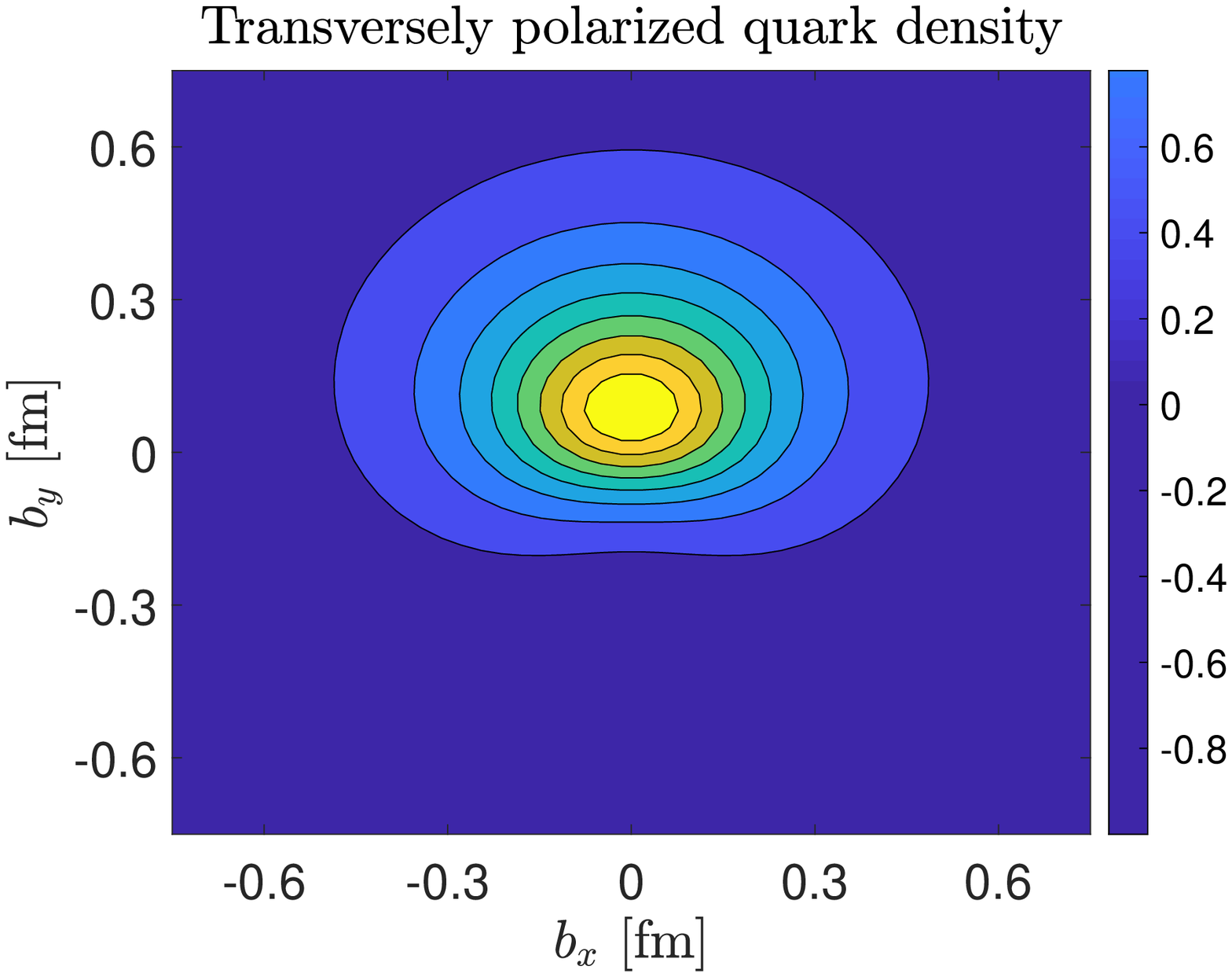}
\caption{Transverse spin densities of pion for unpolarized quark (left panel) and transversely polarized (along $\hat{x}$) quark (right panel).  
}
\label{fig_2}
\end{center}
\end{figure}
\section{Wigner distributions}
Wigner distributions are defined as \cite{Lorce:2011kd}
\begin{equation}
\rho^{[\Gamma]}(x,\bm b_\perp,\bm k_\perp)=\int\frac{\mathrm{d}^2\Delta_\perp}{(2\pi)^2}\,e^{-i\bm\Delta_\perp\cdot\bm b_\perp}\,W^{[\Gamma]}(x,\bm\Delta_\perp,\bm k_\perp),\label{eq:wigner1}
\end{equation}
where $W^{[\Gamma]}(x,\bm\Delta_\perp,\bm k_\perp)$ is the generalized correlator at $\xi=\Delta^+/P^+=0$, and $\bm b_\perp$ is the impact parameter in the position space conjugate to $\bm \Delta_\perp$. Explicitly,
\begin{equation}
W^{[\Gamma]} = \int \frac{dz^- \, d^2 \bm z_\perp}{2 (2\pi)^3} \, e^{i k \cdot z} \,
 \langle p^{\prime} \, | \, \bar{\psi}(-\tfrac{1}{2}z) \,
 \Gamma \, {\cal W} \,
 \psi(\tfrac{1}{2}z) \, | \, p \rangle\, \Big|_{z^+ = 0} \,,
 \label{eq:gcol}
\end{equation}
with $\Gamma\equiv\{\gamma^+, \gamma^{+}\gamma^{5}, i \sigma^{j +}\gamma_5\}$. We define the initial and final momenta of the pion in a symmetric frame as $p'=(p^+, p'^-, \frac{\Delta_\perp}{2})$ and $p''=(p^+, p''^-,- \frac{\Delta_\perp}{2})$, respectively. 
For instance,
\begin{align}
 W^{[\gamma^+]}(x,{\bf \Delta_\perp},{\bm k_\perp})&= \sum_{h', h, \bar{h}} \ \Psi^{*}_{h' \bar{h}}(x,{\bm k''_\perp}) \ \chi^{\dagger}_{h'} \ \chi_{h}\ \Psi_{h \bar{h}}(x,{\bm k'_\perp}), 
\nonumber\\
W^{[\gamma^+ \gamma_5]}(x,{\bf \Delta_\perp},{\bm k_\perp})&= \sum_{h', h, \bar{h}} \ \Psi^{*}_{h' \bar{h}}(x,{\bm k''_\perp})\ \chi^{\dagger}_{h'} \ \sigma_3 \ \chi_{h}\ \Psi_{h \bar{h}}(x,{\bm k'_\perp}),
\nonumber
\\
W^{[i \sigma^{+j} \gamma_5]}(x,{\bf \Delta_\perp},{\bm k_\perp})&= \sum_{h', h, \bar{h}} \ \Psi^{*}_{h' \bar{h}}(x,{\bm k''_\perp}) \ \chi^{\dagger}_{h'}\ \sigma_j \ \chi_{h} \Psi_{h \bar{h}}(x,{\bm k'_\perp}),
\nonumber
\end{align}
where $\sigma_i$ are the Pauli spin matrices and $\chi_{h}$ is the helicity spinor. The arguments
${\bm k'_\perp}$ and ${\bm k''_\perp}$ of the light-front  wavefunctions are given by
$
{\bm k'_\perp}={\bm k}_{\perp} -(1-x){\bm \Delta_\perp\over 2}$,  and ${\bm k''_\perp}={\bm k}_{\perp} +(1-x){\bm \Delta_\perp\over 2}.
$
One then can classify the unpolarized, longitudinally polarized and transversely polarized Wigner distributions in pion as:
$
\rho_{UU}(x,\bm b_\perp,\bm k_\perp)=\rho^{[\gamma^+]}(x,\bm b_\perp,\bm k_\perp),$ 
$\rho_{UL}(x,\bm b_\perp,\bm k_\perp)=\rho^{[\gamma^+\gamma_5]}(x,\bm b_\perp,\bm k_\perp),
$ 
and $\rho_{UT}(x,\bm b_\perp,\bm k_\perp)=\rho^{[i\sigma^{j+}\gamma_5]}(x,\bm b_\perp,\bm k_\perp)$
, respectively.

We compute the pion Wigner distributions using the spin-improved holographic light-front wavefunctions given by~\cite{Ahmady:2018muv} 
	\begin{eqnarray}
	 	\Psi_{h,\bar{h}}(x,\bm{k})= \left[ (M_{\pi} x\bar{x} + B m_f) h\delta_{h,-\bar{h}}  - B    k_\perp e^{-ih\theta_{k_\perp}}\delta_{h,\bar{h}}	\right] \frac{\Psi (x, k_\perp^2)}{x\bar{x}}.
	 \label{spin-improved-wfn-k}
	 \end{eqnarray}
We refer to $B$ as the dynamical spin parameter. $B \to 0$ means no spin-orbit correlations as in the original holographic wavefunction~\cite{Brodsky:2014yha}, while $B \ge 1$ corresponds to a maximal spin-orbit correlations. 
With $B \ge 1$, $m_{u/d}=330$ MeV and a universal AdS/QCD scale, $\kappa=523$ MeV, we successfully predict simultaneously the pion decay constant, charge radius, spacelike electromagnetic and transition form factors, the pion parton distribution functions after taking into account perturbative QCD evolution.
\section{Results}\label{sec3}
In Fig.~\ref{fig_1}, we show the first Mellin moments of Wigner distributions $\rho_{UU}$,
$\rho_{UL}$ and $\rho_{UT}$ for the pion in the upper, central and lower panels. The left panels plot the distributions in the impact-parameter space with fixed transverse momentum $\bm k_\perp = k_\perp \hat{\bm e}_y$ and $k_\perp = 0.3$ GeV, while the right panels plot the distributions in the transverse-momentum space with fixed impact parameter $\bm b_\perp = b_\perp \hat{\bm e}_y$ and $b_\perp = 0.3$ fm. 
We find no distortions in unpolarized quark distributions in both the transverse momentum space and the impact parameter space. They both are circularly symmetric.
However, we observe the dipolar distortion patterns for the longitudinally polarized quark in both spaces and for the transversely polarized quark in impact parameter space only. For the longitudinally polarized quark, the polarity of the impact space distribution is opposite to that in momentum space. When the quark is transversely polarized along $x$-direction, the deformation in $b_\perp$ space appears in $y$-direction.  
These deformation patterns are similar to those of the valence quark distributions of the proton considered as a quark-diquark system~\cite{Chakrabarti:2016yuw}.

Lattice QCD calculations give access
to $x$ moments of quark spin densities. To compare with lattice results, we compute the spin  density for the transversely polarized quark:
\begin{align}
\rho_T({\bm b}_\perp)=\frac{1}{2}\int dx ~d^2 {\bm k}_\perp\,x\,\left[\rho_{UU}(x,\bm b_\perp,\bm k_\perp)+s_T\, \rho_{UT}(x,\bm b_\perp,\bm k_\perp)\right].
\end{align}
Fig.~\ref{fig_2} shows that the unpolarized density is axially symmetric with the peak at the center of pion ($b_\perp=0$), while due to the dipolar distortion from $\rho_{UT}$, the resulting ditribution for a transversely polarized (along +ve $\hat{x}$) quark gets shifted toward positive $\hat{y}$. Our predictions for spin distributions are in qualitative agreement with lattice results~\cite{Brommel:2007xd}.
\\
\\
{\it Acknowledgements:} MA and RS are supported by NSERC (Canada) Grants: SAPIN-2017-00033 and SAPIN-2017-00031, respectively. CM is supported by NSFC (China) under the Grant Nos. 11850410436 and 11950410753. JPV is supported  by the DoE under Grants No.~DE-FG02-87ER40371, and No.~DE-SC0018223 (SciDAC4/NUCLEI). XZ is supported by new faculty startup funding by IMPCAS by Key Research Program of Frontier Sciences, CAS Grant No.~ZDBS-LY-7020.

\end{document}